\documentclass{PoS}

\title{Measurement of the primary Lund jet plane density in pp collisions at 13 TeV with ALICE}

\ShortTitle{ALICE Lund plane}

\author{\speaker{Laura Havener (on behalf of the ALICE collaboration)}\\
        Yale University\\
        E-mail: \email{laura.havener@yale.edu}}

\abstract{  Precision measurements of jet substructure are used as a probe of fundamental QCD processes. The primary Lund jet plane density is a two-dimensional visual representation of the radiation off the primary emitter within the jet that can be used to isolate different regions of the QCD phase space.
A new measurement of the primary Lund plane density for inclusive charged-particle jets in the transverse momentum range of 20 and 120 GeV/$c$ in pp collisions at $\sqrt{s} = $ 13 TeV with the ALICE detector will be presented. This is the first measurement of the Lund plane density in an intermediate jet $p_{\rm T}$ range where hadronization and underlying event effects play a dominant role. The projections of the Lund plane density onto the splitting scale $k_{\rm T}$ and splitting angle $\Delta{R}$ axis are shown, highlighting the perturbative/non-perturbative and wide/narrow angle regions of the splitting phase space. Through a 3D unfolding procedure, the Lund plane density is corrected for detector effects which allows for quantitative comparisons to MC generators to provide insight into how well generators describe different features of the parton shower and hadronization.}

\FullConference{ESP-HEP Conference 2021 \\
26-30 July, 2021 \\
online}

\begin{document}

\section{Introduction}
Jets are a useful probe of fundamental QCD as they undergo many QCD processes during their evolution. Jets originate from  the QCD scattering of partons at high $Q^{2}$. The initial and final partons radiate additional partons which turns into a parton shower that can be described with perturbative QCD (pQCD). Eventually they hadronize, or combine together to form hadrons, which is a non-perturbative QCD (npQCD) process. Finally, there are underlying event effects like multiple parton interactions (MPI) and color neutralization of beam remnants in the same event that are also npQCD processes. All of the final state hadrons from these processes are grouped together to form jets with a defined jet resolution parameter $R$ and transverse momentum $p_{\rm T}$, which are proxies for the initial outgoing partons~\cite{Salam:2010nqg}. Therefore, jets are a useful tool to probe these fundamental QCD processes.


One tool that uses jets to probe QCD is the Lund jet plane density~\cite{Dreyer:2018nbf}, where the Lund plane density is a description of the full phase space of emissions from the jet. The emissions are referred to as jet splittings. Each splitting has two subjets with momenta $p_{\rm T1}$ and $p_{\rm T2}$ that are related by the shared momentum fraction $z = p_{\rm T2}/(p_{\rm T1}+p_{\rm T2})$. In the Lund plane, the $x$-axis is the opening angle between the subjets, $\Delta{R} = \sqrt{\Delta\varphi^{2} + \Delta\eta^{2}}$, where $\Delta\varphi$ and $\Delta\eta$ are the relative azimuthal angle and pseudorapidity between the two subjets. The $y$-axis is the relative transverse momentum (or energy scale) $k_{\rm T} = zp_{\rm T}\sin{\Delta{R}}$, where $p_{\rm T} = p_{\rm T1} + p_{\rm T2}$. From here it follows that diagonal lines along the plane represent constant $z$ values. The presented analysis deals with the \emph{primary Lund jet plane density}.
This represents a Lund plane which is filled 
with all the splitting that can be found along the trajectory that follows the harder branch in each splitting
when unwinding the clusterization history. The density in the plane at leading order is described as 
\begin{equation}
\rho(\Delta{R},k_{\rm T})= \frac{1}{N^{\rm{jets}}}\frac{\rm{d}^{2}\it{n}}{\rm{d}\ln({\it{R}}/\Delta \it{R})\rm{d}\ln(\it{k_{\rm T}})},
\label{eq:Lund}
\end{equation}
where $N^{\rm{jets}}$ is the total number of jets and $n$ is the total number of splittings in the selected jet $p_{\rm T}$ interval~\cite{Dreyer:2018nbf}. At leading order the emissions populate the Lund plane uniformly. It is interesting to look at deviations from uniformity like how the running of the coupling sculpts the shape of the plane. 

The Lund diagram can then be used to isolate different regions of QCD phase space. First, the $k_{\rm T}$ axis can separate the perturbative from non-perturbative dominated regimes. Additionally, diagonal $z$ lines separate harder splittings from softer splittings. The different QCD processes fall in different regions of the plane, for example underlying event effects are located at small $k_{\rm T}$ and large angles, perturbative splittings are at high $k_{\rm T}$  and large angles, and hadronization effects are at low $k_{\rm T}$ . This separation allows for detailed comparisons to MC generators to see how well the models are describing the different underlying processes. 

The ATLAS collaboration recently measured the primary Lund jet plane density for jets at very high $p_{\rm T}$ ($ > 675$ GeV/$c$)~\cite{Aad:2020zcn}. This measurement was fully corrected to particle level using a three-dimensional unfolding procedure. Therefore, comparisons could be made to MC generators to constrain these generators at high $p_{\rm T}$. ALICE is well-suited to perform jet substructure measurement at more intermediate values of the jet $p_{\rm T}$ due to high precision tracking in the Time Projection Chamber (TPC) and Inner Tracking System (ITS). Therefore, ALICE can perform a complementary measurement to ATLAS in a different kinematic regime where non-perturbative effects like hadronization and the underlying event play a more dominant role. In these proceedings, a new measurement of the primary Lund jet plane density in pp collision at $\sqrt{s} = $ 13 TeV from the ALICE experiment will be shown compared to various MC generators~\cite{ALICE-PUBLIC-2021-002}. 

\section{Analysis method}

\subsection{Jet reconstruction}
The jets are reconstructed using the anti-$k_{\rm T}$ algorithm \cite{FastJetAntikt} and the $E$-scheme \cite{Cacciari:2011ma} recombination with resolution parameter $R=0.4$, assuming the mass of the pion for each track. The required minimum track $p_{\rm T}$ is 150 MeV$/c$ and the track acceptance is $|\eta| < 0.9$ and $0 < \varphi < 2\pi$. The jet $p_{\rm T}$ range is between 20 and 120 GeV/$c$ and the jet axis must be inside the fiducial acceptance of the TPC, which means at least one jet radius away from the edges of the TPC ($-0.5 < \eta < 0.5$ and $0 < \varphi < 2\pi$). Jets with a track with $p_{\rm T}$ above 100 GeV$/c$ are not used in the analysis due to poor momentum resolution of tracks in that $p_{\rm T}$ range, however the fraction of such jets in the considered jet momentum range is negligible. 

\subsection{Lund plane}
Once the jets are reconstructed, their constituents are re-clustered with the C/A algorithm to enforce angular ordering~\cite{FastJetCA}. The clustering sequence is undone and the branches from the clustering are followed from the leading prong. At each clustering step  two subjet prongs $p_{\rm{T},1}$, $p_{\rm{T},2}$ with $p_{\rm{T},1} \geq p_{\rm{T},2}$ are found and the Lund plane is filled with the kinematics $\ln{(k_{\rm{T}})}$ and $\ln{(R/\Delta{R})}$ of the subleading prong. 

\subsection{Unfolding}
\begin{figure}[ht!]
    \centering
    \includegraphics[scale = 0.35]{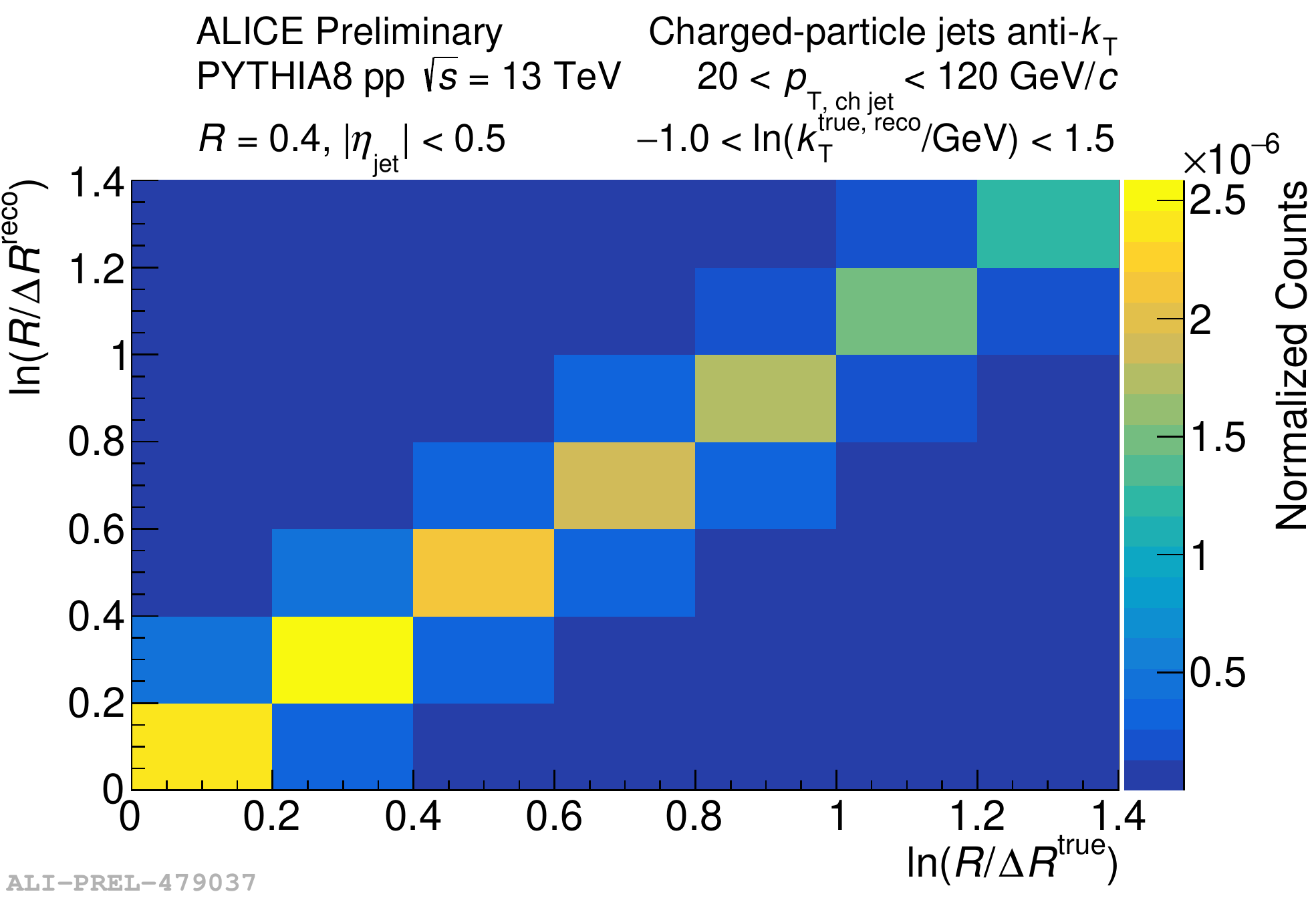}
    \includegraphics[scale = 0.35]{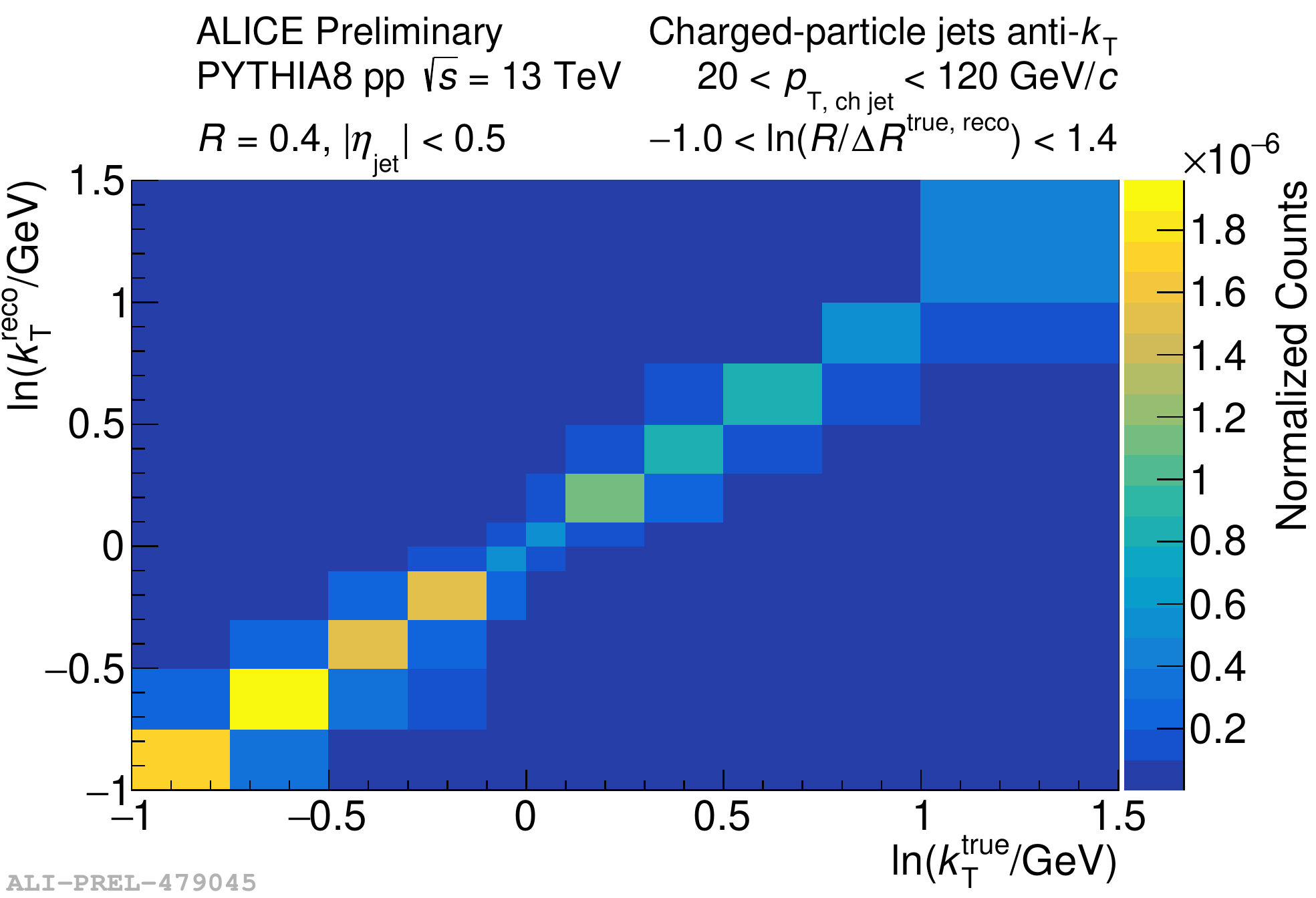} 
    \caption{Detector response of the inverse splitting angle $R/\Delta{R}$ and splitting scale $k_{\rm T}$. True and reco labels denote generator and detector-level, respectively.}
    \label{fig:Resp}
\end{figure}

This measurement requires a 3D unfolding for detector effects at the level of the individual splittings in the jet $p_{\rm T}$ and the two axes of the Lund plane. The response is built by matching individual splittings between the true jets from Pythia 8~\cite{Sjostrand:2007gs} pp simulations and the reconstructed jets that were ran through a GEANT3~\cite{GEANT3} ALICE detector simulation. The detector response for the angular and the $k_{\rm T}$ scale are shown in Figure~\ref{fig:Resp}. Both responses are fairly diagonal indicating great resolution for the splittings which allows the measurement to be unfolded for detector effects.

\begin{figure}[ht!]
    \centering
    \includegraphics[scale = 0.35]{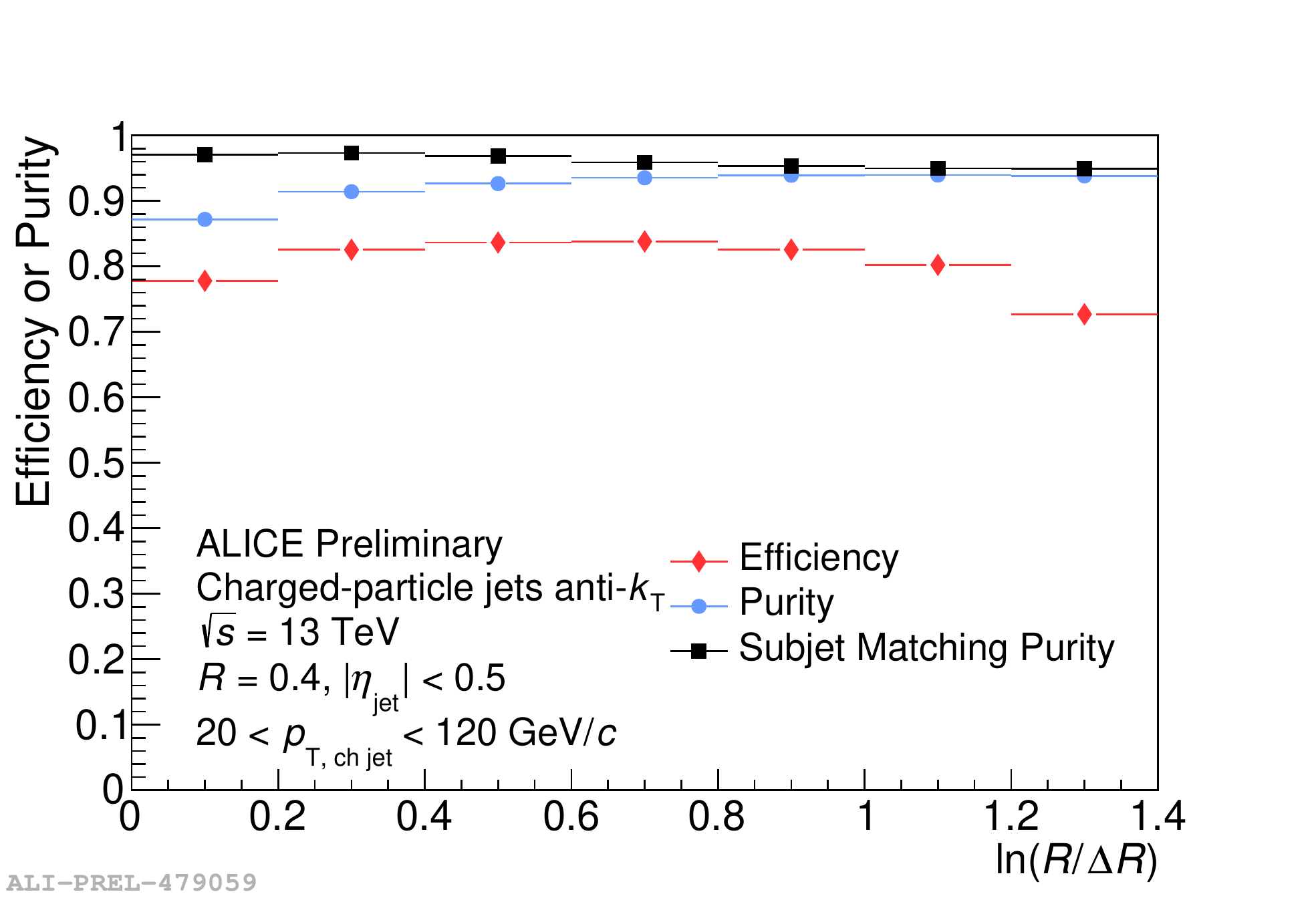}
    \includegraphics[scale = 0.35]{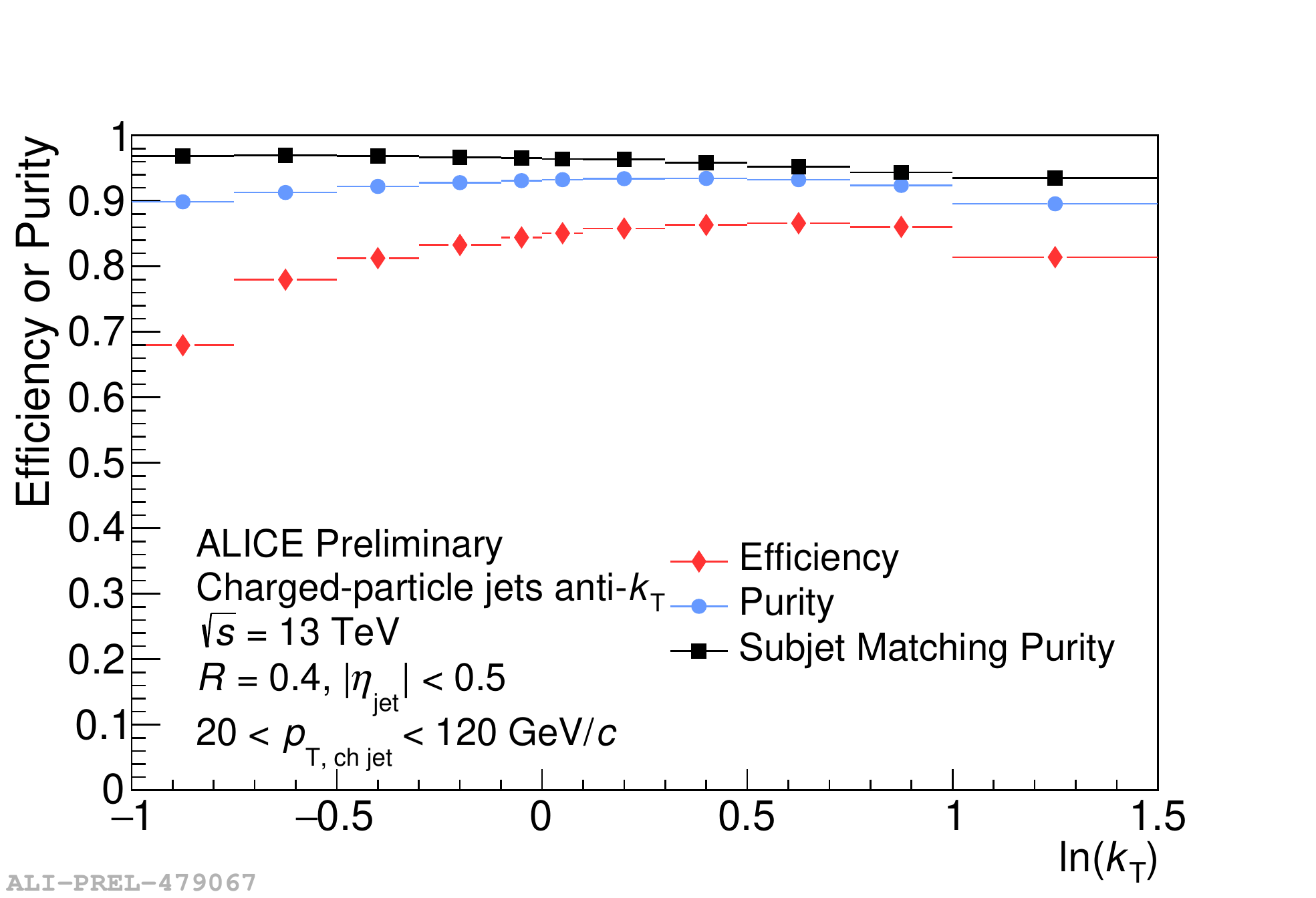} 
    \caption{The efficiency and purity in red and blue, respectively, for the splitting angle (left) and splitting $k_{\rm{T}}$ (right). The fraction of correct matches is also shown in black.}
    \label{fig:Eff}
\end{figure}

The analysis procedure requires a detailed study of the efficiency and purity of the jet splittings. The efficiency is defined as $e=N_{\rm true}^{\rm match}/N_{\rm true}$, where $N_{\rm true}$ is the number of true splittings and $N_{\rm true}^{\rm match}$ is the number of matched true splittings. The purity is defined as $p=N_{\rm reco}^{\rm match}/N_{\rm reco}$, where $N_{\rm reco}$ is the number of reconstructed splittings and $N_{\rm reco}^{\rm match}$ is the number of matched reconstructive splittings. The efficiency and purity are shown in Figure~\ref{fig:Eff}, where the left panel is the $\ln{(R/\Delta{R})}$ axis and the right panel is the $\ln{(k_{\rm{T}})}$ axis. Both are shown to be above 80\%, indicating high efficiency and purity. In the unfolding procedure the raw data is corrected for the purity before unfolding. After unfolding the distribution is corrected for the efficiency. Finally, the Lund plane is normalized using a 1D unfolded jet spectra to get a per jet normalization. 

Additionally, the subjet matching purity is evaluated which determines if the true splittings were correctly identified in the reconstructed splitting. This is quantified by evaluating the fraction of momentum of the true splitting carried by the corresponding tracks in the geometrically matched reconstructed splitting. When the fraction of momentum is greater than or equal to $50\%$, the pair is called a correct match. If the shared momentum is less than $50\%$, the pair is called a mismatch. The subjet purity is very high ($ > 90\%$), see the black points in Figure~\ref{fig:Eff}.

\section{Results} 
\begin{figure}[ht!]
    \centering
    \includegraphics[scale = 0.5]{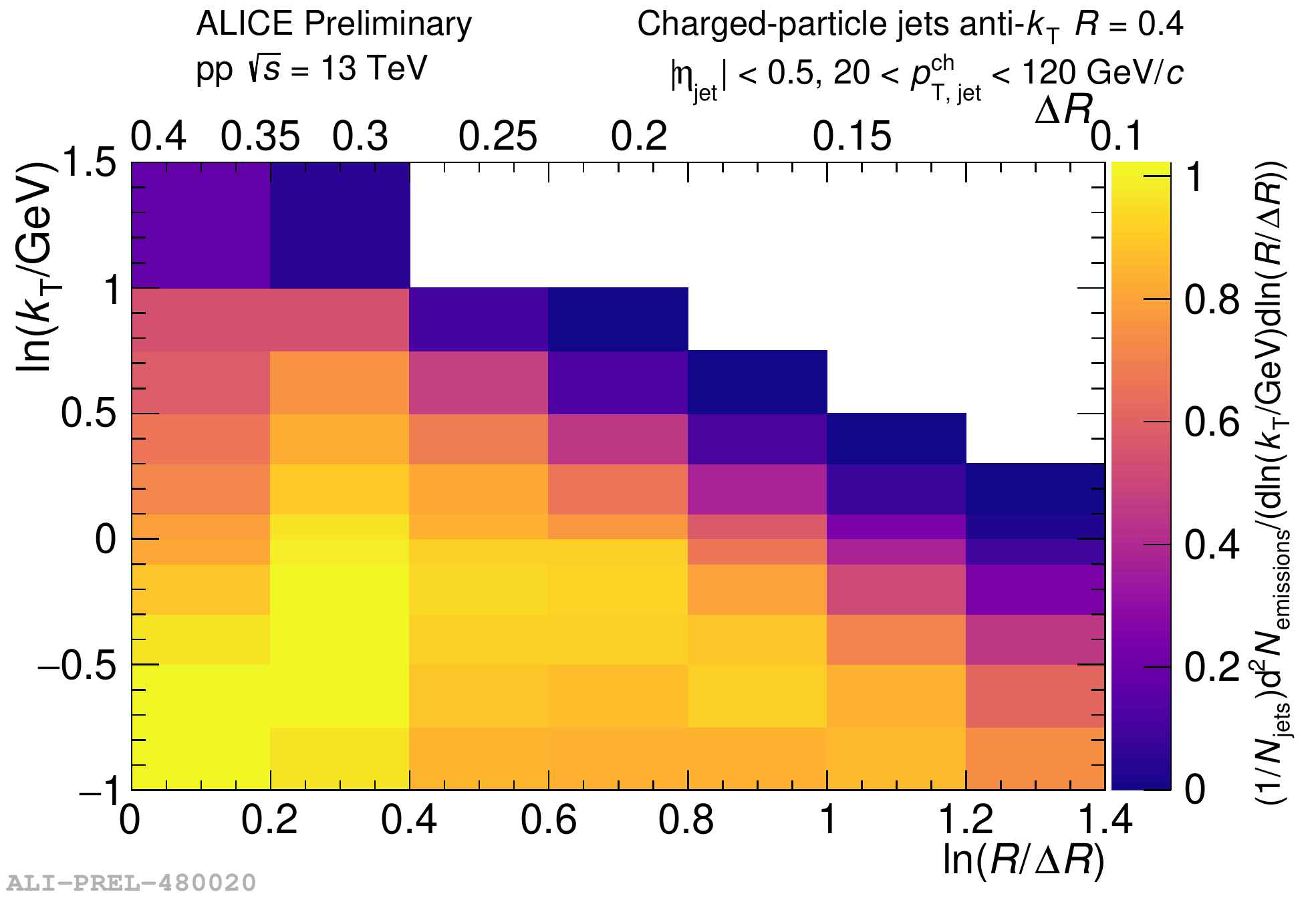}
    \caption{Fully corrected primary Lund plane density.}
    \label{fig:Lund}
\end{figure}

Final unfolded results are presented and compared to different MC generators. First, the fully corrected Lund jet plane density in pp collisions for charged jets between 20 and 120 GeV/$c$ is shown in Figure~\ref{fig:Lund}. Jet splittings were measured out to a $k_{\rm T}$ of 5 GeV/$c$ and for angular distances between 0.1 and 0.4. It is interesting to make projections in this plane to isolate regions of QCD phase space and make detailed comparisons to MC generators. 

The different MC generators used in this analysis have different implementations of the parton shower and hadronization. Four generators are used: PYTHIA8 Monash~\cite{Sjostrand:2007gs}, Herwig 7~\cite{Bellm:2015jjp}, and Sherpa 2.2.8~\cite{Gleisberg:2008ta} with two settings. For the parton shower implementation, PYTHIA is $k_{\rm T}$ ordered, Herwig is angular ordered, and Sherpa uses a dipole shower ordering. For the hadronzation, PYTHIA and the Lund Sherpa setting use a Lund string hadronization and Herwig and the AHADIC Sherpa setting use a cluster hadronization. 

In this analysis, various systematic uncertainties were considered including uncertainties on the tracking efficiency, unfolding, and model dependence. The dominating uncertainty is the model dependence which considers the dependence of the detector effects on the fragmentation model. This is evaluated by using both Herwig 7 and PYTHIA8 fast detector simulations for the unfolding and corrections and comparing the final unfolded results. This uncertainty can be as large as 10\% at large angles or high $k_{\rm T}$. 

\begin{figure}[ht!]
    \centering
    \includegraphics[scale = 0.35]{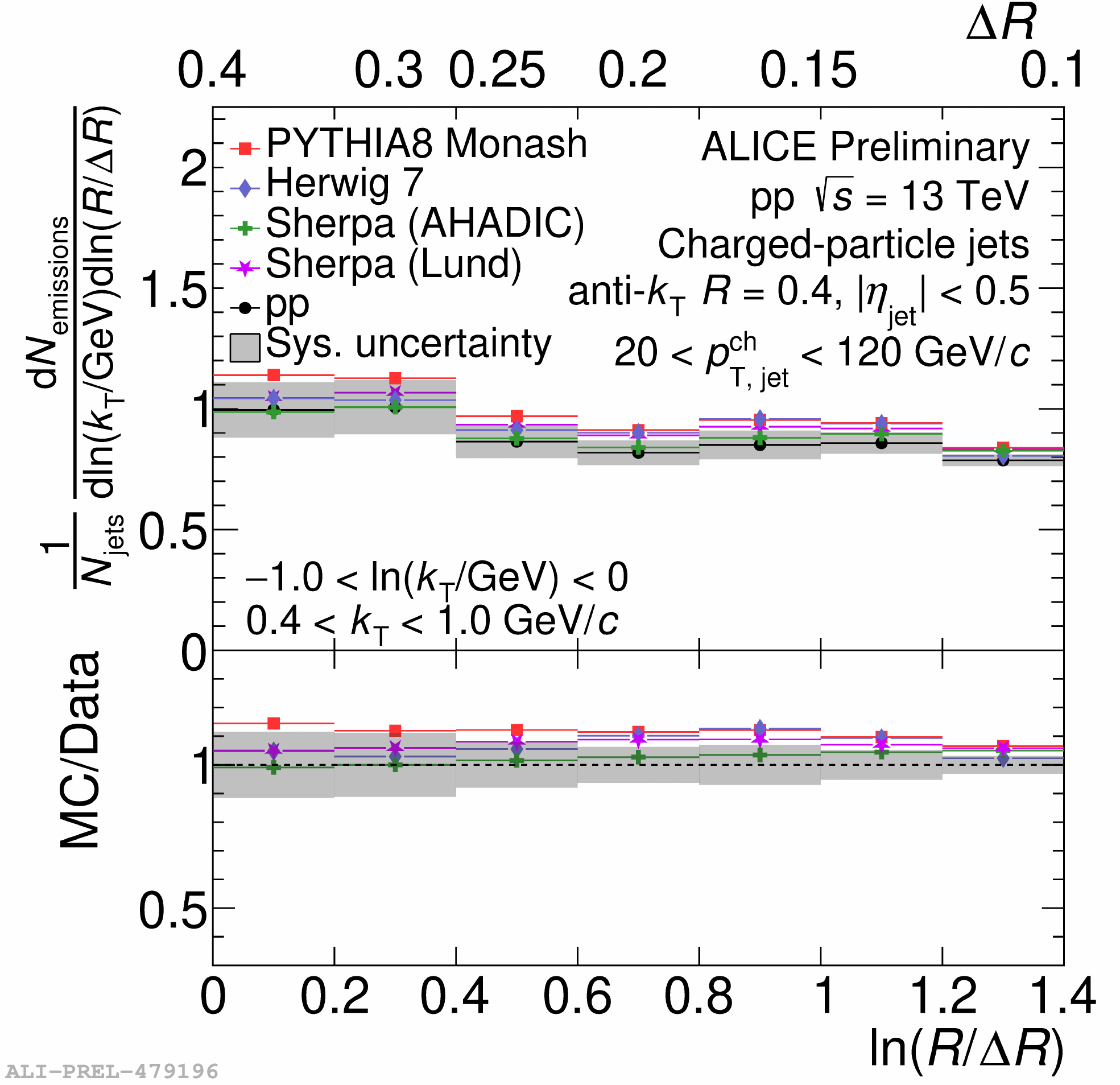}
    \includegraphics[scale = 0.35]{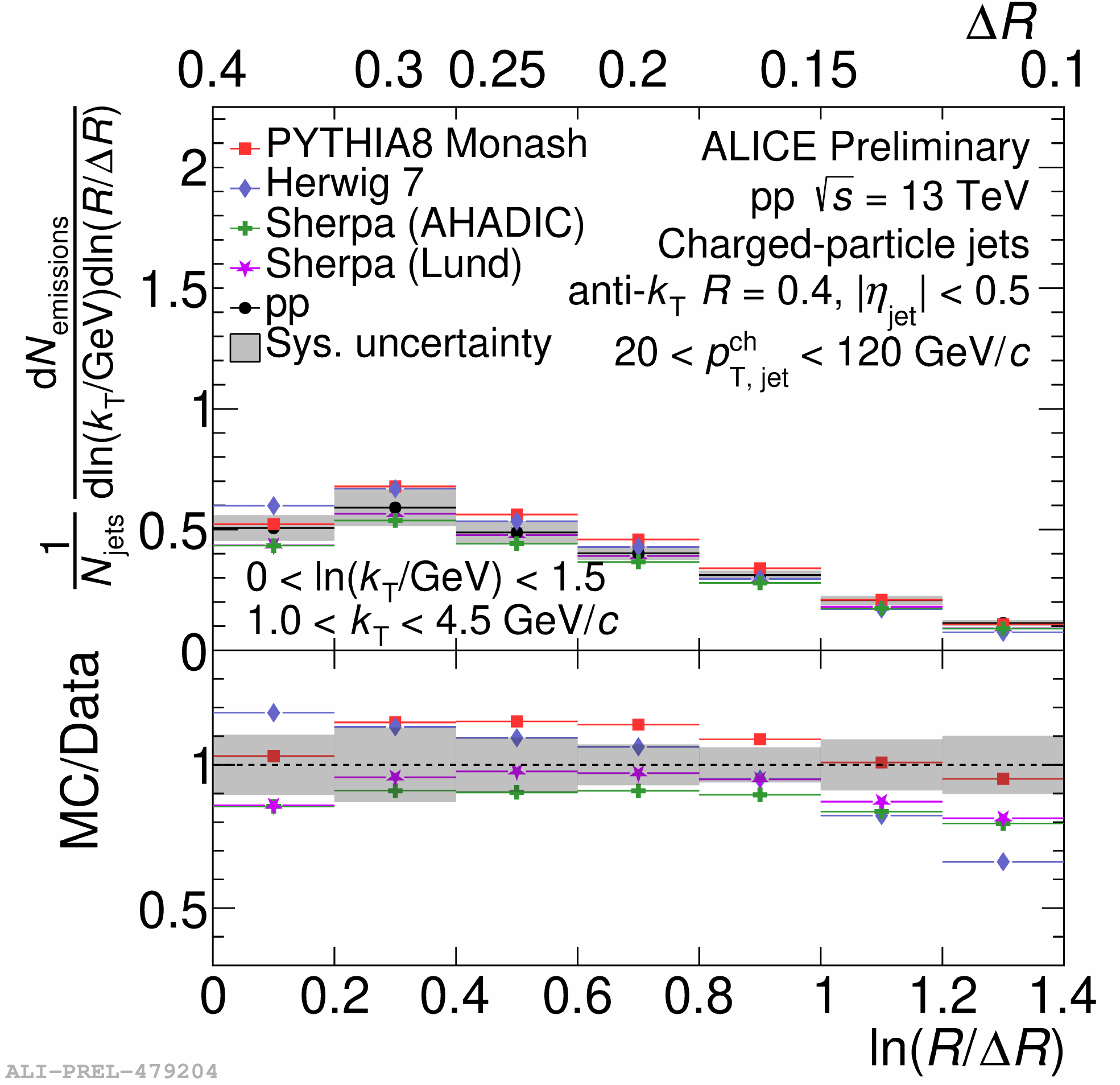}  \\
    \includegraphics[scale = 0.35]{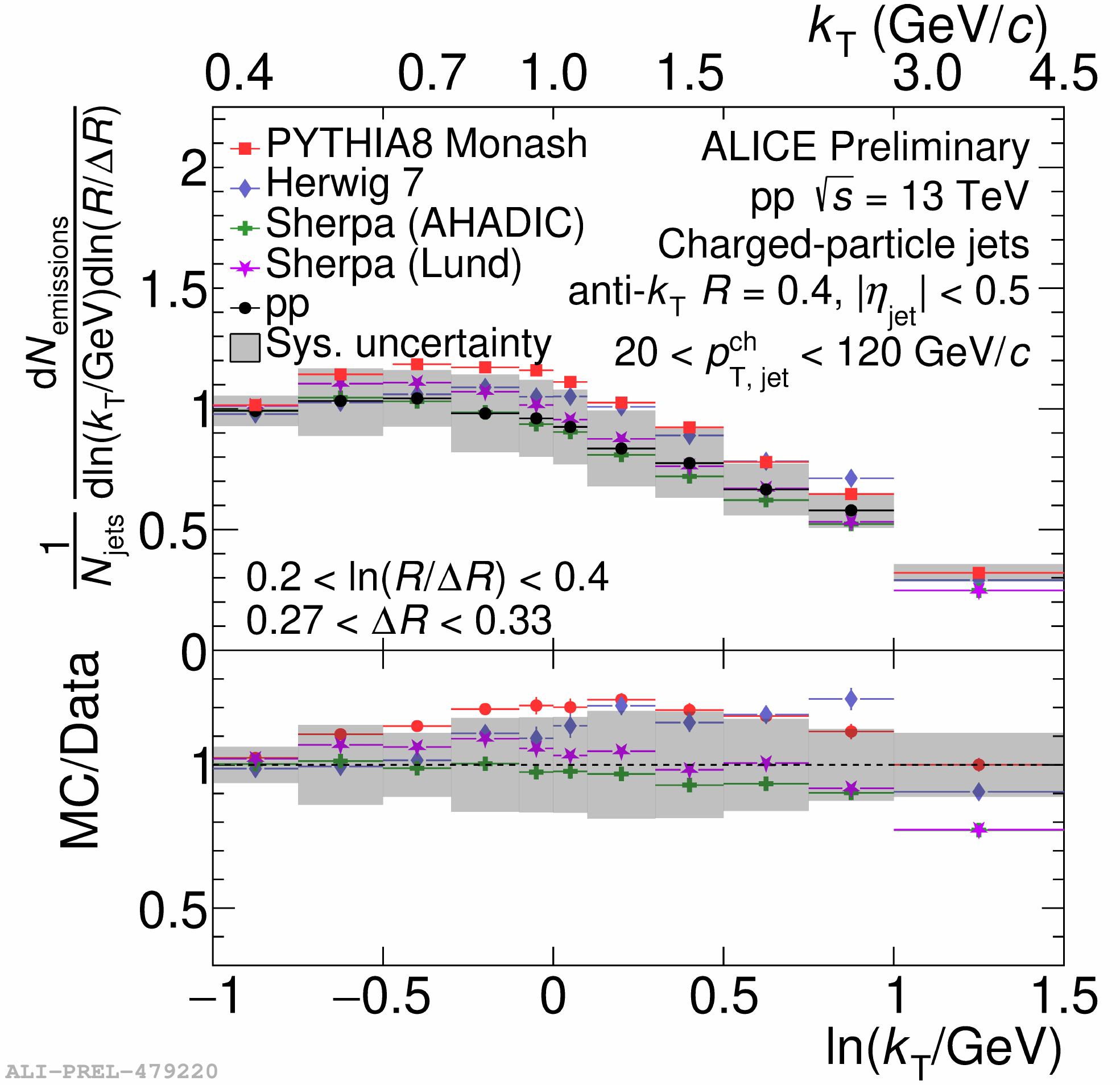}
    \includegraphics[scale = 0.35]{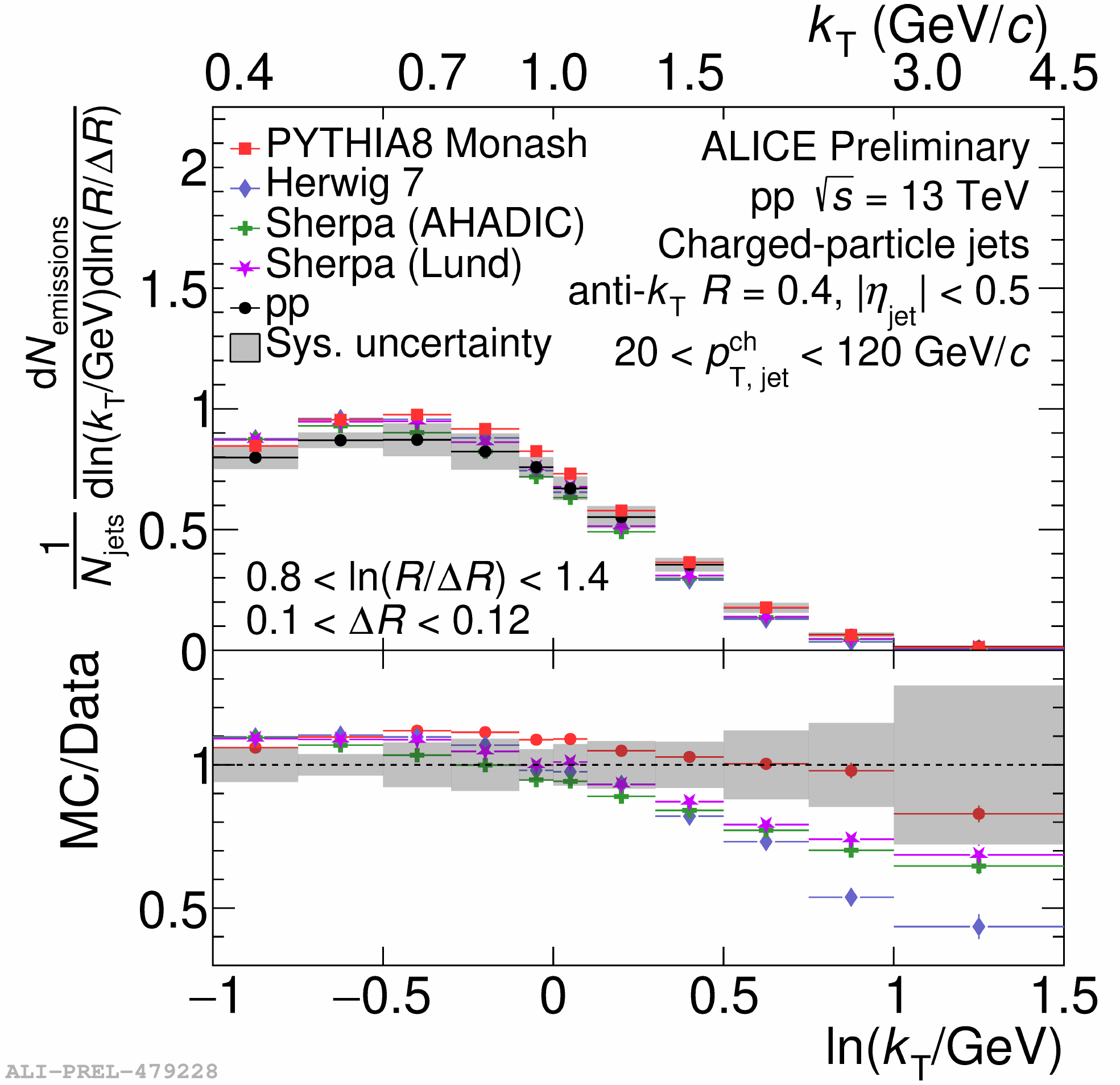}
    \caption{The projections of the primary Lund jet plane density onto the $\ln({R/\Delta R)}$ (top) and $\ln{(k_{T})}$ (bottom) axes compared to different MC generators for different regions of the Lund plane. The top left shows the angular dependence for a non-perturbative region and the top right for a perturbative region. The bottom panels show the $k_{\rm T}$ distribution for wider splittings on the left and narrower splittings on the right. The ratios of the generators to the data are shown in the bottom panel. }
    \label{fig:Proj}
\end{figure}

Projections onto the angular axis are shown in Figure~\ref{fig:Proj}. The right panel corresponds to more perturbative splittings ($\ln{(k_{\rm{T}})} > 0$) and the left panel presents more non-perturbative splittings ($\ln{(k_{\rm{T}})} < 0$). The bottom panel shows the ratio of the various MC generators to the data where the MC describes the data to within 10\% in most cases. In the high-$k_{\rm{T}}$ region, the MC falls below the data for some of the models, particularly Herwig. 

Projections onto the $k_{\rm{T}}$ axis are shown in Figure~\ref{fig:Proj}. The left panel corresponds to wider splittings ($0.2 < \ln{(R/\Delta{R})} < 0.4$) and the right panel to narrower splittings ($0.8 < \ln{(R/\Delta{R})} < 1.4$). Again, there is agreement within 10\% for most of the MC generators. There is a significant difference between data and MC at high $k_{\rm{T}}$ for collinear splittings, in particular for Herwig.

\section{Conclusion}
The fully corrected measurement of the primary Lund jet plane density has been presented for charged-particle jets in pp collisions at $\sqrt{s}=13$ TeV with the ALICE detector. This is the first measurement of the primary Lund plane density in an intermediate jet $p_{\rm T}$ range where hadronization and underlying event effects play a dominant role.  A 3D unfolding procedure was used to correct the Lund plane density for detector effects, allowing for quantitative comparisons to MC generators. Projections of the Lund plane density are shown in order to isolate different QCD phase space regions and provide insight into how well generators describe different underlying QCD processes like the parton shower and hadronization. The data seems to be described by the generators within 10\% except for the most collinear, highest-$k_{\rm T}$ splittings where Herwig underestimates the data. 
The next step is to apply this method in future jet-triggered data samples in ALICE to increase the $k_{\rm T}$ reach and compare to analytical calculations.

\bibliographystyle{JHEP}
\bibliography{mybib}

\providecommand{\href}[2]{#2}\begingroup\raggedright\begin{thebibliography}{10}

\bibitem{Salam:2010nqg}
G.P.~Salam, \emph{{Towards Jetography}},
  \href{https://doi.org/10.1140/epjc/s10052-010-1314-6}{\emph{Eur. Phys. J. C}
  {\bfseries 67} (2010) 637} [\href{https://arxiv.org/abs/0906.1833}{{\ttfamily
  0906.1833}}].

\bibitem{Dreyer:2018nbf}
F.A.~Dreyer, G.P.~Salam and G.~Soyez, \emph{{The Lund Jet Plane}},
  \href{https://doi.org/10.1007/JHEP12(2018)064}{\emph{JHEP} {\bfseries 12}
  (2018) 064}.

\bibitem{Aad:2020zcn}
{\scshape ATLAS} collaboration, \emph{{Measurement of the Lund Jet Plane Using
  Charged Particles in 13 TeV Proton-Proton Collisions with the ATLAS
  Detector}}, \href{https://doi.org/10.1103/PhysRevLett.124.222002}{\emph{Phys.
  Rev. Lett.} {\bfseries 124} (2020) 222002}
  [\href{https://arxiv.org/abs/2004.03540}{{\ttfamily 2004.03540}}].

\bibitem{ALICE-PUBLIC-2021-002}
{\scshape ALICE} collaboration, \emph{{Physics Preliminary Summary: Measurement
  of the primary Lund plane density in pp collisions at $\sqrt{s} = \rm{13}$
  TeV with ALICE}},  Mar, 2021.

\bibitem{FastJetAntikt}
M.~Cacciari, G.P.~Salam and G.~Soyez, \emph{{The anti-$k_t$ jet clustering
  algorithm}}, \href{https://doi.org/10.1088/1126-6708/2008/04/063}{\emph{JHEP}
  {\bfseries 04} (2008) 063}.

\bibitem{Cacciari:2011ma}
M.~Cacciari, G.P.~Salam and G.~Soyez, \emph{{FastJet User Manual}},
  \href{https://doi.org/10.1140/epjc/s10052-012-1896-2}{\emph{Eur. Phys. J. C}
  {\bfseries 72} (2012) 1896}
  [\href{https://arxiv.org/abs/1111.6097}{{\ttfamily 1111.6097}}].

\bibitem{FastJetCA}
Y.L.~Dokshitzer, G.D.~Leder, S.~Moretti and B.R.Webber, \emph{{Better Jet
  Clustering Algorithms}},
  \href{https://doi.org/10.1088/1126-6708/1997/08/001}{\emph{JHEP} {\bfseries
  08} (1997) 001}.

\bibitem{Sjostrand:2007gs}
T.~Sjostrand, S.~Mrenna and P.Z.~Skands, \emph{{A Brief Introduction to PYTHIA
  8.1}}, \href{https://doi.org/10.1016/j.cpc.2008.01.036}{\emph{Comput. Phys.
  Commun.} {\bfseries 178} (2008) 852}
  [\href{https://arxiv.org/abs/0710.3820}{{\ttfamily 0710.3820}}].

\bibitem{GEANT3}
{R. Brun, F. Bruyant, M. Maire, A.C. McPherson, and P. Zanarini}, \emph{{GEANT3
  User's Guide}}, {\emph{CERN Data Handling Division DD/EE/84-1} (1985) }.

\bibitem{Bellm:2015jjp}
J.~Bellm et~al., \emph{{Herwig 7.0/Herwig++ 3.0 release note}},
  \href{https://doi.org/10.1140/epjc/s10052-016-4018-8}{\emph{Eur. Phys. J. C}
  {\bfseries 76} (2016) 196}
  [\href{https://arxiv.org/abs/1512.01178}{{\ttfamily 1512.01178}}].

\bibitem{Gleisberg:2008ta}
T.~Gleisberg, S.~Hoeche, F.~Krauss, M.~Schonherr, S.~Schumann, F.~Siegert
  et~al., \emph{{Event generation with SHERPA 1.1}},
  \href{https://doi.org/10.1088/1126-6708/2009/02/007}{\emph{JHEP} {\bfseries
  02} (2009) 007} [\href{https://arxiv.org/abs/0811.4622}{{\ttfamily
  0811.4622}}].

\end{thebibliography}\endgroup



\end{document}